\documentclass{article}
\usepackage{amssymb}

\usepackage{graphicx}
\usepackage{amsmath}
\usepackage{mathrsfs}
\usepackage{textcomp}

\begin{document}

\title{Feynman, Wigner, and Hamiltonian Structures Describing the Dynamics of Open
Quantum Systems}

\author{J. Gough\footnote{Institute of Mathematics, Physics, and Computer Sciences, Aberystwyth
University, Aberystwyth, United Kingdom, email: \texttt{jug@aber.ac.uk}}, 
T.S. Ratiu\footnote{Section de Math\'{e}matiques and Bernoulli Center, \'{E}cole Polytechnique
F\'{e}d\'{e}rale de Lausanne, Lausanne, CH 1015 Switzerland, email: \texttt{tudor.ratiu@epfl.ch}}, 
O.G. Smolyanov\footnote{Mechanics and Mathematics Faculty, Moscow State University, Moscow, 119991
Russia, 
email: \texttt{smolyanov@yandex.ru}}}

\date{}

\maketitle

\begin{abstract}
This paper discusses several methods for describing the dynamics of open quantum systems, where the environment of the open
system is infinite-dimensional.  These are purifications, phase space forms, master equation and liouville equation forms.
The main contribution is in using Feynman-Kac formalisms to describe the infinite-demsional components.
\end{abstract}

This paper discusses several approaches for describing the dynamics of open 
quantum systems. Open quantum systems play an important role in modelling physical 
systems coupled to their environment and, in particular, for the emerging field of quantum
feedback control theory (see [12]). Thus, in studying coherent quantum feedback,
models consisting of a quantum system related to quantum control system, so
that each of these systems turns out to be open, are considered.

Generally the master equation is presented in the theoretical physics literature as the 
central description of an open quantum system. In practice, however, it is these solutions that
are important for potential applications, rather than the master equations themselves. 
Our goal is to solve the exact master equations, which describe the
reduced dynamics of subsystems of certain large systems generated by the
dynamics of these large systems: these master equation arise from a number of different 
approaches which we will consider.

In fact, we examine four approaches for describing
subsystems dynamics, and in each case we exploit Feynman type formulas (see [1,
2]). Moreover, we assume that the quantum systems under consideration are
obtained by the Schr\"{o}dinger quantization [3] of classical Hamiltonian
systems.

\begin{enumerate}
\item
Our first approach is based on a representation of mixed states as
random pure ones, where the dynamics of a subsystem of the isolated
quantum system is described by a random
process taking values in the Hilbert space of the subsystem. The random
process is then defined by using the Feynman formula for the solution of the
Schr\"{o}dinger equation for the united system.

\item
The second approach uses the
Wigner function [4] and also its infinite dimensional analogue, the Wigner
measure, which was introduced in [5]. If the phase space of the classical
Hamiltonian system generating the quantum system under consideration is
finite dimensional, then the density of the Wigner measure with respect to
the standard Lebesgue measure coincides with the Wigner function. As shown
in [5], the evolution of the Wigner measure of a closed quantum system is
described by a Liouville-Moyal type equation; in order to obtain a solution of the
master equation for the dynamics of the Wigner measure (or function) of a
subsystem of the initial system from a solution of this equation represented
by using a Feynman type formula, it suffices to integrate this representation
over the coordinates of the phase space of the corresponding classical
subsystem. Another approach for describing the evolution of the Wigner
function of a subsystem is discussed in [6].

\item
The third approach again uses the
Feynman formulas for the Schr\"{o}dinger equation for a quantum system
and its environment, but this time these formulas are used to
describe the evolution of the density operator of each part; it is given by
the corresponding partial trace of the evolving density operator of the
united system.

\item
The final approach considered here is based on the representation of any state of the
quantum system by a probability measure on its Hilbert space. In the case of
a closed system, the evolution of this measure is described by the
Liouville equation generated by the Hamiltonian structure, and here the
Hamiltonian equation coincides with the Schr\"{o}dinger equation (see [7-9]).
At the same time, the correlation operator of this measure coincides with
the density operator [10], so knowing the evolution of the density
operator of the subsystem allows us to obtain the evolution of the probability
measure on its Hilbert space and hence solve the master
equation generated by the associated Liouville equation. It should perhaps be emphasized 
that the technique for treating Wigner measures by employing a 
suitable projection of the (pseudo)measure defined on the
space of the combined classical system, is not applicable in this situation
 because the Hilbert space of the united system is the tensor product,
rather than the Cartesian product.
\end{enumerate}

This paper focuses on the algebraic structures related to the problems
under consideration, and we do not explicitly state the analytical
details.

\section{STATES OF OPEN QUANTUM SYSTEMS}

Let $\mathfrak{H}_{1}$ and $\mathfrak{H}_{2}$ be the Hilbert spaces of states of two
quantum systems. In what follows, we refer to the first system (as well as
to its classical counterpart) as the \emph{open system} and to the second as \emph{the
environment}, respectively. These two systems form a composite system, whose Hilbert space
is the Hilbert tensor product [3] $\mathfrak{H}=\mathfrak{H}_{1}\otimes \mathfrak{H}_{2}$.

Let $\mathscr{Q}_{1}$ and $\mathscr{Q}_{2}$ be the configuration spaces of the corresponding
classical open system and  the environment, respectively. We assume that $\mathscr{Q}_{1}$
and $\mathscr{Q}_{2}$ are real separable Hilbert spaces. In both cases we shall
assume that we have measures $\nu_j$ ($j=1,2$) defined on the $\sigma$-algebra of Borel subsets of the
corresponding spaces. We
then set
$$\mathfrak{H}_{1}=L^{2}(\mathscr{Q}_{1},\nu_1) , \quad \mathfrak{H}%
_{2}=L^{2}(\mathscr{Q}_{2},\nu_2)$$
and in particular the composite Hilbert space is then
$$\mathfrak{H}=\mathfrak{H}%
_{1}\otimes \mathfrak{H}_{2} \cong L^{2}(\mathscr{Q}_{1}\times \mathscr{Q}_{2},\nu
_{L}\otimes \nu_2).$$

The open system we wish to
describe will be quantum mechanical, so we have dim$\, \mathscr{Q}
_{1}<\infty $, and  fix  $\nu_1$ to be standard Lebesgue measure on $\mathscr{Q}
_{1}$ for definiteness.

The dimension of $\mathscr{Q}_{2}$ will typically be infinite, in this case,
according to the well-known result of Weil, there does not exist a non-zero $\sigma $-finite
countably additive locally finite Borel measure on $\mathscr{Q}_{2}$.
Instead, we fix a Gaussian measure $\nu_2$ on $\mathscr{Q}%
_{2}$ (this is a matter of convenience, however, and non-Gaussian measures may be used as well). 

If $\varphi \in L^{2}(\mathscr{Q}_{1}\times \mathscr{Q}_{2},\nu_1\otimes
\nu_2)$ is normalized, that is, 
$$\int_{\mathscr{Q}_{1}\times \mathscr{Q}%
_{2}}\left| \varphi \left( q_{1},q_{2}\right) \right| ^{2} \nu_1(dq_{1})\nu_2(dq_{2})=1,$$
then the \emph{marginal distributions} $\rho _{k}$
are defined by 
\begin{eqnarray*}
\rho _{1}(q_{1})\triangleq \int_{\mathscr{Q}_{2}}\left| \varphi \left( q_{1},q_{2}\right) \right| ^{2}\nu_2(dq_{2}) \\
\rho _{2}(q_{2})\triangleq \int_{\mathscr{Q}_{1}}\left| \varphi \left( q_{1},q_{2}\right) \right| ^{2}\nu_1(dq_{1})
\end{eqnarray*}
A probability measure $\mathbb{P}_{2}$ on $\mathscr{Q}_{2}$ is then defined by
\begin{eqnarray*}
\mathbb{P}_{2 } (dq_2)= \rho_2 (q_2 ) \, \nu_2 (dq_2 ),
\end{eqnarray*}
so $\mathbb{P}_2$ is absolutely continuous with respect $\nu_2$ with Radon-Nikodym density $\rho_2$, and describes the results of
measurements of the environment coordinates. The pair $(\mathscr{Q}_{2},\mathbb{P}_{2})$ is a
Kolmogorov probability space. 

Taking the fixed pure state $\varphi$, we may define a $\mathfrak{H}_1 (=L^{2}(\mathscr{Q}_{1},\nu_1))$-valued random variable on
$(\mathscr{Q}_{2},\mathbb{P}_{2})$ by
\begin{eqnarray}
\Psi_1:  q_{2}\mapsto \varphi
\left( \cdot ,q_{2}\right)  .
\label{eq:F}
\end{eqnarray}

If we fix an operator $\hat{A}_1$ on $\mathfrak{H}_1$, then
\begin{eqnarray*}
\mathbb{E}_2 \left[ \frac{ \langle \Psi_1
\vert \hat{A}_1 \vert \Psi_1 \rangle_{\mathfrak{H}_1} }
{\langle \Psi_1
\vert  \Psi_1 \rangle_{\mathfrak{H}_1}} \right]
&=&   \int_{\mathscr{Q}_2}  \frac{ \langle \Psi_1
\vert \hat{A}_1 \vert \Psi_1 \rangle_{\mathfrak{H}_1} (q_2) }
{\langle \Psi_1
\vert  \Psi_1 \rangle_{\mathfrak{H}_1} (q_2) }
\, \mathbb{P}_2 (dq_2) 
\\
&=&   \int_{\mathscr{Q}_2}  \frac{ \int_{\mathscr{Q}_1} \varphi (q_1 ,q_2 )^\ast (\hat{A}_1 \otimes I_2 \, \varphi ) (q_1,q_2) \nu_1 (dq_1)}
{\int_{\mathscr{Q}_1} \vert \varphi (q^\prime_1 ,q_2 ) \vert^ 2 \nu_1 (dq^\prime_1) } \mathbb{P}_2 (dq_2) \\
&=&  \int_{\mathscr{Q}_1} \int_{\mathscr{Q}_2}   \varphi (q_1 ,q_2 )^\ast (\hat{A}_1 \otimes I_2 \, \varphi ) (q_1,q_2) \nu_1 (dq_1)
\nu_2 (dq_2) \\
&\equiv& \langle \varphi \vert \hat{A}_1 \otimes I_2 \vert \varphi \rangle_{\mathfrak{H}_1 \otimes \mathfrak{H}_2} \\
& \equiv & \mathrm{tr}_{\mathfrak{H}_1} [ \hat{\varrho}_1 \hat{A}_1 ].
\end{eqnarray*}
where $\hat{\varrho}_1$ is the von Neumann density operator corresponding the the marginal state of the open system.

\textbf{Proposition 1}. \textit{The correlation operator of the probability
measure on }$L^{2}(\mathscr{Q}_{1},\nu_1)$\textit{, which is the distribution of
results of measurements of the random pure state } $\Psi_1$ \textit{ given in (\ref{eq:F}),
coincides with the von Neumann density operator}.

In Dirac notation, we may write $\ langle q_1 \vert \Psi \rangle $ for the complex-valued random variable $\langle q_1 \vert \Psi : q_2 \mapsto \varphi (q_1 , q_2 )$, then
\begin{eqnarray*}
\mathbb{E}_2 \left[ \frac{ \langle q_1 \vert \Psi_1 \rangle 
\langle \Psi_1  \vert q^\prime_1 \rangle  }
{\langle \Psi_1
\vert  \Psi_1 \rangle_{\mathfrak{H}_1}} \right]
= \varrho( q_1 , q^\prime_1 )
,
\end{eqnarray*}
where $\varrho( q_1 , q^\prime_1 )$ is the kernel operator of $\hat{\varrho}_1$.

\bigskip

\textbf{Remark 1}. It is useful to compare the following two approaches for
calculating the probability distribution of the results of measurements of
the coordinate $q_{1}$ of the first system, one of which directly uses the
function $\varphi \in L^{2}(\mathscr{Q}_{1}\times \mathscr{Q}_{2},\nu_1
\otimes \nu_2)$, which represents the pure state of the composite system,
and the other one uses the $L^{2}(\mathscr{Q}_{1},\nu_1)$-valued random variable 
$\Psi_1 $. In the former case, the marginal probability density $\rho _{1}$ giving the results
of measurements of the coordinate $q_{1}$.
In the second approach, the density $\rho _{1}$ can be obtained by using
the Chapman-Kolmogorov formula, the random variable $F$, and the probability 
$\mathbb{P}_{2}$ as 
\begin{eqnarray*}
\rho _{1}(q_{1}) &= &\int_{\mathscr{Q}_{2}}\rho _{1}(q_{1}|q_{2})%
\mathbb{P}_{2}(dq_{2}) \\
&=&\int_{\mathscr{Q}_{2}}\frac{\left| \varphi \left( q_{1},q_{2}\right)
\right| ^{2}}{\int_{\mathscr{Q}_{1}}\left| \varphi \left( q_{1}^{\prime
},q_{2}\right) \right| ^{2}\nu_1(dq_{1}^{\prime })}\int_{\mathscr{Q}%
_{1}}\left| \varphi \left( q_{1}^{\prime \prime },q_{2}\right) \right|
^{2}\nu_1(dq_{1}^{\prime \prime })\nu_2(dq_{2}) \\
&\equiv &\int_{\mathscr{Q}_{2}}\left| \varphi \left( q_{1},q_{2}\right)
\right| ^{2}\nu_2(dq_{2}),
\end{eqnarray*}
where the conditional probability density $\rho _{1}(q_{1}|q_{2})$ is
defined by $\rho _{1}(q_{1}|q_{2})=\left| \varphi \left( q_{1},q_{2}\right)
\right| ^{2}/\int_{\mathscr{Q}_{1}}\left| \varphi \left( q_{1}^{\prime
},q_{2}\right) \right| ^{2}\nu_1(dq_{1}^{\prime })$.

\bigskip

\textbf{Remark 2}. Of course, the Hilbert-valued random variable representing
a mixed state of the open system is not uniquely determined; e.g., instead
of the \textit{coordinate representation} $L^{2}(\mathscr{Q}_{1},\nu_1)$
of the Hilbert space $\mathfrak{H}_{1}$, we can use a \textit{momentum
representation} of this Hilbert space.

\bigskip

\textbf{Remark 3}. It also follows from the above considerations that the
evolution of the open system can be described by a random process taking
value in the same Hilbert space. However, this process is not uniquely
determined either; thus, the corresponding master equation (which is an
equation with a time dependent random coefficient) is not uniquely
determined.

\section{RANDOM PROCESSES DESCRIBING \\
THE EVOLUTION OF OPEN SYSTEMS}

We recall that the Feynman formulas are representations of
Schr\"{o}dinger groups or semigroups as limits of integrals over finite
Cartesian products of some space $\mathscr{X}$, (see, e.g., [1, 2]). If $\mathscr{X}$ coincides with the domain $%
\Omega $ of functions from the space on which these groups or semigroups act
and $\Omega \subset \mathscr{Q}=\mathscr{Q}_{1}\times \mathscr{Q}_{2}$ then
the corresponding Feynman formula is said to be \textit{Lagrangian}; if $%
\mathscr{X}=\Omega \times \mathscr{P}$, where $\mathscr{P}=\mathscr{P}_{1}\times %
\mathscr{P}_{2}$ is the momentum space of the classical version of the
quantum system under consideration, then the Feynman formula is said to be
Hamiltonian (not all Feynman formulas belong to one of these two classes)%
\footnote{The Feynman-Kac formulas are representations of the same groups and
semigroups as integrals over the space consisting of functions of a real
variable taking values in the same space $\mathscr{X}$. The multiple integrals in the
Feynman formulas approximate the (infinite dimensional) integrals in the
Feynman-Kac formulas. In the case of the Schr\"{o}dinger semigroups generated
by Hamiltonians quadratic in momenta, the infinite dimensional integrals in
the Feynman-Kac formulas turn out to be integrals with respect to probability
measures; however, in the case of Schr\"{o}dinger groups, there appear
integrals with respect to the so-called Feynman pseudo-measures or their
analogues in the Feynman-Kac formulas (in many realistic situations,
integrals with respect to pseudo-measures are defined as the limits of
appropriate sequences of finitely many integrals).}.

We identify (see [4, 10]) $\mathscr{P}_{j}$ with $\mathscr{Q}_{j}^{\ast }$%
and $\mathscr{Q}_{j}$ with $\mathscr{P}_{j}^{\ast }$ $(j=1,2)$. These
identifications generate isomorphisms (cf. [4]) 
\begin{equation}
J:\mathscr{Q}_{j}\times \mathscr{P}_{j}\ni \left( q,p\right) \mapsto \left(
p,q\right) \in \left( \mathscr{Q}_{j}\times \mathscr{P}_{j}\right) ^{\ast }
\end{equation}
and a similar isomorphism between the spaces 
\begin{equation}
\mathscr{Q}\times \mathscr{P}\triangleq \left( \mathscr{Q}_{1}\times %
\mathscr{Q}_{2}\right) \times \left( \mathscr{P}_{1}\times \mathscr{P}%
_{2}\right)
\end{equation}
and $\left( \mathscr{Q}\times \mathscr{P}\right) ^{\ast }$.

Let $\psi _{1}\in \mathfrak{H}_{1}\left( =L^{2}(\mathscr{Q}_{1},\nu_1\right) $
be the initial state of the open system, and let $\psi _{2}$ be the initial
state of the environment, which is called the \textit{reference state}. We
have $\left( \psi _{1}\otimes \psi _{2}\right) \left( q_{1},q_{2}\right)
=\psi _{1}(q_{1})\psi _{2}(q_{2})$. Suppose that a classical Hamiltonian
function $H:\mathscr{Q}\times \mathscr{P}\mapsto \mathbb{R}$ is defined by 
\begin{equation}
H\left( q_{1},p_{1},q_{2},p_{2}\right) \triangleq H_{1}\left(
q_{1},p_{1}\right) +H_{2}\left( q_{2},p_{2}\right) +H_{12}\left(
q_{1},p_{1},q_{2},p_{2}\right)
\end{equation}

The  Hamiltonian observable $\hat{H}$ governing the
evolution of the composite system may be written as 
\begin{equation}
\hat{H}=\hat{H}_{1}\otimes I_{2}+I_{1}\otimes \hat{H}_{2}+\hat{H}_{12},
\end{equation}
where $\hat{H}_{j}$ is the pseudo-differential operator on $\mathfrak{H}_{j}$
with Weyl symbol $H_{j}$ for $j=1,2$ and $\hat{H}_{12}$ is the
pseudo-differential operator on with Weyl symbol $H_{12}$ (for the definition
of pseudo-differential operators on spaces of functions square integrable
with respect to a measure different from the Lebesgue measure, see [2, 10]).
It is useful to assume that $\hat{H}_{1}$ governs the internal dynamics of
the open system, $\hat{H}_{2}$ governs the internal dynamics of the
environment, and $\hat{H}_{12}$ describes the interaction.

\bigskip

\textbf{Theorem 1}. \textit{Suppose that, for each }$t=0$\textit{, }$\varphi
(t)\in H_{1}\otimes H_{2}$\textit{\ denotes the state of the composite
system at the moment }$t$\textit{. Then, for all }$(q_{1},q_{2})\in %
\mathscr{Q}_{1}\times \mathscr{Q}_{2}$\textit{,} 
\begin{eqnarray*}
\varphi (t)\left( q_{1},q_{2}\right) &=&\left( e^{it\hat{H}}\,\psi
_{1}\otimes \psi _{2}\right) \left( q_{1},q_{2}\right) =\lim_{n\rightarrow
\infty }\left( \widehat{e^{i\frac{t}{n}H}}\right) ^{n}\,\psi _{1}\otimes
\psi _{2}\left( q_{1},q_{2}\right) \\
&=&\lim_{n\rightarrow \infty }\left( \widehat{e^{i\frac{t}{n}H_{1}\otimes
I_{2}}}\circ \widehat{e^{i\frac{t}{n}I_{1}\otimes H}}\circ \widehat{e^{i%
\frac{t}{n}H_{12}}}\right) ^{n}\,\psi _{1}\otimes \psi _{2}\left(
q_{1},q_{2}\right) .
\end{eqnarray*}

The proof is based on Chernoff's theorem [11].

\bigskip

\textbf{Remark 4}. The substitution of the explicit expressions for the
pseudo-differential operators on the right-hand side of the last relation
turns this relation into a Feynman type formula.

\bigskip

We now define two random processes describing the dynamics of the open
quantum system. Suppose that, for each $t\geq 0$, $\mathbb{P}_{t}$ is a
probability measure on a copy $\mathscr{Q}_{2}^{t}$ of the space $\mathscr{Q}%
_{2}$ whose density $\rho _{t}\left( \cdot \right) $ with respect to $\nu_2$
is defined as 
\begin{equation}
\rho _{t}(q_{2})\triangleq \int_{\mathscr{Q}_{1}}\left| \lim_{n\rightarrow
\infty }\left( \widehat{e^{i\frac{t}{n}H}}\right) ^{n}\,\psi _{1}\otimes
\psi _{2}\left( q_{1},q_{2}\right) \right|^2 \nu_1(dq_{1}),
\end{equation}
$\mathbb{P}$ is the probability measure on the product space $\mathscr{X}$ of the
family of spaces $\left\{ \mathscr{Q}_{2}^{t}:t\geq 0\right\} $ defined as
the product of the measures $\mathbb{P}_{t}$, and $\psi ^{\mathbb{P}%
}:[0,\infty )\times \left( \mathscr{X},\mathbb{P}\right) \mapsto L^{2}(%
\mathscr{Q}_{1})$ is the $L^{2}(\mathscr{Q}_{1})$-valued random process
defined by 
\begin{equation}
\psi ^{\mathbb{P}}\left( t,q\right) =\psi _{t}^{\mathbb{P}}(q)\triangleq
\varphi \left( t\right) \left( \cdot ,q(t)\right)
\end{equation}
where $q(=q(\cdot ))\in \mathscr{X}$ and $\varphi $ is the pure state function 
appearing in 
Theorem 1. Suppose also that, for the same $t$, $\gamma (t)$ is a bijection
between $\mathscr{Q}_{2}$ and $\mathscr{Q}_{2}^{t}$, which determines
an isomorphism between the measure space $(\mathscr{Q}_{2},\nu_2)$ and the
measure space $(\mathscr{Q}_{2}^{t},\mathbb{P}_{t})$, and $\psi
_{v}:[0,\infty )\times \left( \mathscr{Q}_{2},\nu_2\right) \mapsto L^{2}(%
\mathscr{Q}_{1}, \nu_1 )$ is the random process defined by 
\begin{eqnarray}
\psi _{\nu
}(t,q)\triangleq \varphi \left( t\right) \left( \cdot ,\gamma (t)\left(
q\right) \right) .
\end{eqnarray}

\bigskip

\textbf{Theorem 2}. \textit{Under the above assumptions, the state of
the open quantum system at a moment of time }$t$\textit{\ is described by the }$L_{2}(\mathscr{Q}_{1}, \nu_1)$\textit{-valued random
variables} $\psi^{\mathbb{P}} (t, \cdot )$ \textit{ (on } $(\mathscr{X},\mathbb{P})$ \textit{) and } $\psi_{\nu_2}(t,q)$
\textit{ (on (}$\mathscr{Q}_{2},\nu_2)$ \textit{).}

\section{THE WIGNER EVOLUTION FUNCTIONS OF THE OPEN QUANTUM SYSTEM}

Given a density operator $T$ on $\mathfrak{H}$, the Weyl function generated by $%
T $ is the function $W_{T}:\mathscr{Q}\times \mathscr{P}\mapsto \mathbb{R}$
defined by 
\begin{equation}
W_{T}(H)\triangleq \mathrm{tr}\left\{ Te^{-i\hat{H}}\right\} ,
\end{equation}
where $\hat{H}$\ is the pseudo-differential operator on $\mathfrak{H}=L^{2}(%
\mathscr{Q},\nu_1\otimes \nu_2)$ with symbol $JH\in \mathscr{Q}^{\ast
}\times \mathscr{P}^{\ast }$ [5]. The \textit{Wigner measure} on $\mathscr{Q}%
\times \mathscr{P}$ generated by the density operator $T$ is defined by 
\begin{equation}
\int_{\mathscr{Q}\times \mathscr{P}}e^{i\left( \mathfrak{p}_{1}\mathfrak{q}_{2}-%
\mathfrak{q}_{1}\mathfrak{p}_{2}\right) }W_{T}^{M}\left( d\mathfrak{q}_{1},d\mathfrak{p}%
_{1}\right) =W_{T}\left( \mathfrak{q}_{1},\mathfrak{p}_{2}\right) ,
\end{equation}
with $\left( \mathfrak{q},\mathfrak{p}\right) \in \mathscr{Q}\times \mathscr{P}$, cf. [5].

The Wigner measure $W_{T_{1}}^{M}$ on $\mathscr{Q}_{1}\times \mathscr{P}_{1}$
generated by a density operator $T_{1}$ on $\mathfrak{H}_{1}$ is defined in a
similar way. The density of the measure $W_{T_{1}}^{M}$\ with respect to $%
\nu_1$ coincides with the classical Wigner function (see [5]).

\bigskip

\textbf{Theorem 3}. \textit{If }$T$\textit{\ is a density operator on }$H$%
\textit{\ and }$T_{1}$\textit{\ is the corresponding reduced density
operator on }$H_{1}$\textit{, then} 
\begin{equation*}
W_{T_{1}}^{M}\left( \cdot \right) =\int_{\mathscr{Q}_{2}\times \mathscr{P}%
_{2}}W_{T}^{M}\left( \cdot ,dq_{2},dp_{2}\right) .
\end{equation*}

Using this theorem and the Feynman formula for the solution of the Moyal
type equation which describes the evolution of the Wigner measure on $%
\mathscr{Q}\times \mathscr{P}$, we can obtain a formula describing the
evolution of the Wigner measure (and, thereby, the Wigner function) on $%
\mathscr{Q}_{1}\times \ \mathscr{P}_{1}$.

\section{HAMILTONIAN STRUCTURES}

This section considers the third and the fourth approach for describing the
dynamics of open quantum systems, which are closely related to each other.
We assume that $\psi _{1}$, $\psi _{2}$, and $\mathfrak{H}$\ are the same as
above and $T(\cdot )$ is a function describing the dynamics of the open
system, whose values are density operators on $\mathfrak{H}_{1}$.

\bigskip

\textbf{Theorem 4}. \textit{If, for each }$t>0$\textit{, }$k_{T}(t)$\textit{%
\ is the integral kernel of a trace-class operator }$T(t)$\textit{\ on }$%
H_{1} $\textit{, then} 
\begin{eqnarray*}
k_{T}\left( t,q_{1},q_{2}\right) &=&\int_{\mathscr{Q}_{2}}\left[
\lim_{n\rightarrow \infty }\left( \widehat{e^{i\frac{t}{n}H}}\right)
^{n}\,\psi _{1}\otimes \psi _{2}\left( q_{1},q\right) \right] \\
&& \times \left[
\lim_{n\rightarrow \infty }\left( \widehat{e^{i\frac{t}{n}H}}\right)
^{n}\,\psi _{1}\otimes \psi _{2}\left( q_{2},q\right) \right] \nu_2(dq).
\end{eqnarray*}

\bigskip

\textbf{Theorem 5}. \textit{Let }$\nu_2(\cdot )$\textit{\ be a function of a
real variable such that, for each }$t$\textit{, }$\nu_{2,t}$\textit{\ is a
Gaussian measure on }$\mathfrak{H}_{1}$\textit{\ with correlation operator }$T(t)$%
\textit{. Then the function }$\nu_{2,t}$\textit{\ satisfies the master
(Liouville) equation.}

\bigskip

\textit{ACKNOWLEDGMENTS}

This work was supported by the Government of Russian Federation, state
contract no. 11.G34.31.0054. T.S. Ratiu acknowledges the support of the
Switzerland National Scientific Foundation, Swiss NSF grant no.
200021140238. O.G. Smolyanov acknowledges the support of his visit to
Aberystwyth by London Mathematical Society.

\end{document}